\documentclass[twocolumn]{jpsj3}
\usepackage{txfonts}
\usepackage{bm}
\usepackage{color}
\title{Relationship between Ferromagnetic Criticality and the Enhancement of Superconductivity Induced by Transverse Magnetic Fields in UCoGe}

\author{Taisuke Hattori$^1$\thanks{E-mail: t.hattori@scphys.kyoto-u.ac.jp},
Kosuke Karube$^1$,
Kenji Ishida$^1$\thanks{E-mail: kishida@scphys.kyoto-u.ac.jp},
Kazuhiko Deguchi$^2$,
Noriaki K. Sato$^2$,
and Tomoo Yamamura$^3$}

\inst{$^1$Department of Physics, Graduate School of Science, Kyoto University, Kyoto 606-8502, Japan \\
$^2$Department of Physics, Graduate School of Science, Nagoya University, Nagoya 464-8602, Japan\\
$^3$Institute for Materials Research, Tohoku University, Sendai 980-8577, Japan} 

\abst{We have performed \Co~NMR experiments on the ferromagnetic superconductor UCoGe under magnetic fields ($H$) along the $a$- and $b$- axes to investigate the relationship between ferromagnetic properties and superconductivity. The ferromagnetic ordering temperature \TCurie~is suppressed and the nuclear spin-lattice relaxation rate $1/T_1$ at 2 K is enhanced in $H~||~b$, although \TCurie~and $1/T_1$ are unchanged in $H~|| ~a$, indicating that the ferromagnetic criticality is induced only when $H$ is applied along the $b$ axis. We show the close relationship between the magnetic anisotropies and the superconducting ones reported by Aoki {\it et al.}: the superconductivity is gradually suppressed in $H~||~a$, but enhanced in $H~||~b$ above 5 T.
We strongly suggest that the enhancement of the superconductivity observed in $H~||~b$ originates from the field induced ferromagnetic criticality, as pointed out by Aoki {\it et al} and Mineev.
}

\newcommand{\Co}{$^{59}$Co}

\newcommand{\TCurie}{\ensuremath{T_{\rm Curie}}}
\newcommand{\TSC}{\ensuremath{T_{\rm SC}}}
\newcommand{\Hc}{\ensuremath{H_{\rm c2}}}

\begin{document}
\maketitle
UCoGe exhibits the ferromagnetic (FM) ordering at a low Curie temperature $\TCurie$ of $\sim 3$ K and a superconducting (SC) transition temperature $\TSC$ of $\sim 0.8$ K at ambient pressure\cite{NTHuy2007}, which is the highest among the FM superconductors discovered so far. Although UCoGe possesses a three-dimensional crystal structure, its magnetic property is the Ising anisotropy with the $c$-axis being the easy axis\cite{NTHuy2008,DAoki2009JPSJ}. In addition, its SC upper critical fields (\Hc) also have strong anisotropy; superconductivity survives far beyond the Pauli-limiting field along the $a$- and $b$- axes, whereas \Hc~along the $c$-axis is as small as 0.5 T \cite{NTHuy2008,DAoki2009JPSJ}.
 
Since UCoGe includes the familiar NMR-active nucleus \Co, it is a suitable compound for NMR measurements within the FM superconductors. We have shown that superconductivity occurs in the FM region from the \Co~nuclear quadrupole resonance (NQR), resulting in the microscopic coexistence of ferromagnetism and superconductivity. This is consistent with the $\mu$SR result\cite{ADevisser2009}. From the precise angle-resolved NMR at 1.7 K and Meissner measurements at 85 mK, we show that the magnetic field along the $c$- axis strongly suppresses the Ising FM fluctuations along the $c$- axis and that the superconductivity is observed only in the limited magnetic field region where the FM spin fluctuations are active\cite{THattori2012PRL}. These results, combined with model calculations, strongly suggest that the Ising FM spin fluctuations tuned by $H~||~c$ induce the unique spin-triplet superconductivity by resolving the above puzzling \Hc~behavior.

Although the external magnetic field along the $c$-axis is the tuning parameter of the Ising FM fluctuations and the key to understanding the small $H_{\rm c2}$ along the $c$-axis, the magnetic field along the $b$-axis, which is the second magnetic easy axis and perpendicular to the U-U zigzag chain, also tunes the superconductivity \cite{DAoki2009JPSJ}. 
The superconductivity becomes robust against the external field along the $b$ axis when $\mu_0 H$ greater than 5 T is applied, which is reminiscent of the re-entrant superconductivity in the sister compound URhGe \cite{FLevy2005URhGeReSC, FLevy2007URhGePD}, as discussed later. 
However, such a robustness of the superconductivity was not observed in $H~||~a$.
Although $T_{\rm SC}$ suppression by $H^{a}$ and $H^{b}$ is the same in the field smaller than $\mu_{0}H^{a,b} <$ 4 T, the different response of the superconductivity against the field greater than 5 T along the $a$- and $b$-axes would give another important clue for understanding the superconductivity in UCoGe. 
From the macroscopic viewpoint, these increases in \TSC~in URhGe and UCoGe are suggested to originate from the increase in the effective mass at approximately the FM critical point \cite{AMiyake2008JPSJ, DAoki2009JPSJ}. To understand the origin of the anisotropy of superconductivity as well as the increased effective masses in $H~||~b$, we have investigated the anisotropy of magnetic properties directly by NMR measurements in the fields along the $a$- and $b$-axes, since NMR is a powerful experimental technique for probing low-energy spin dynamics.

\begin{figure}[htbp]
 \includegraphics[width=\hsize,clip]{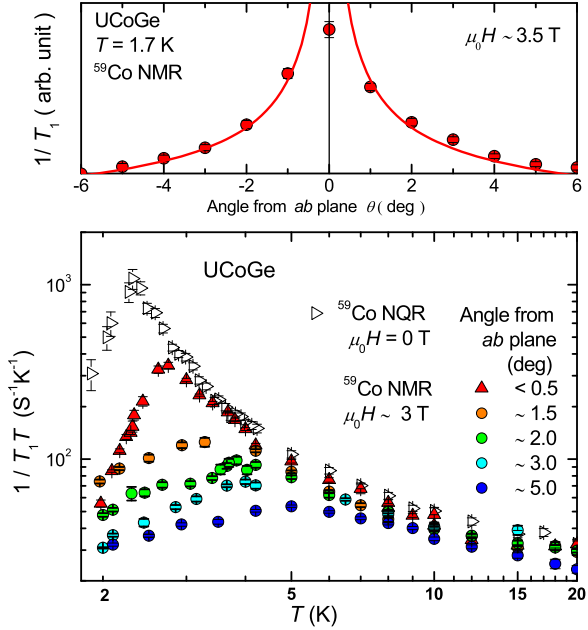}
 \vspace*{-10pt}
 \caption{(color online) Angular dependence of $1/T_1$ in the $bc$ plane; the line is a guide to the eye (upper figure\cite{THattori2012PRL}). Temperature dependence of $1/T_1T$ measured by \Co~NMR under various angles $\theta$ (bottom figure). $\theta$ is the angle between the field direction and the $ab$ plane. \Co-NQR $1/T_1T$ of UCoGe is also shown as a zero-field result.}
  \label{Fig:T1T-HcDep}
\end{figure}

The single-crystalline UCoGe grown by the Czochralski crystal pulling method in a tetra-arc furnace under high-purity argon was utilized for the measurement, which is the same sample reported previously\cite{TOhta2010JPSJ, YIhara2010PRL, THattori2012PRL}. 
The sample shows a relatively large residual resistivity ratio of approximately 30 along the $b$-axis. The FM transition temperature \TCurie~was evaluated to be $2.55 \pm 0.1$ K from the Arrot plots, and the midpoint SC transition temperature was determined from the ac susceptibility as 0.57 K.  
\Co~NMR measurements were done in the longitudinal 15 T SC magnet using a double-axis rotator mounted in the NMR probe. 
$1/T_1$ was determined by fitting the recovery curves $R(t)$ of the nuclear magnetization $m(t)$ at $t$ after the saturation pulse to the theoretical function. 
The field direction to the crystal axis is carefully checked using the simulation of the \Co~NMR peak locus, as described in a previous paper\cite{THattori2011JPSJS}. 
In addition, we also took advantage of the very high sensitivity of the $1/T_1$ value against the $c$-axis component of the external field, as shown in the upper figure of Fig.~\ref{Fig:T1T-HcDep}. $1/T_1T$ is proportional to the magnetic fluctuations perpendicular to the field direction when the Zeeman energy attributable to the applied field is much larger than that attributalbe to the electric quadrupole interaction at the Co site (in the case of UCoGe, $\mu_0 H \gg 0.28$ T).

As discussed in the previous paper\cite{THattori2012PRL}, the divergence of $1/T_1T$ at $\TCurie$ of $\sim 2.5$ K in the field along the $a$-and $b$-axes indicates that the Ising FM fluctuations along the $c$-axis become critical. Therefore, the sensitivity of $1/T_1T$ against the field along the $c$ axis ($H^c$) implies the significant suppression of the Ising FM fluctuations with $H^c$. 
As increasing $H^c$ by rotating the single-crystal sample, the peak temperature of $1/T_1T$ is increased continuously, accompanied by the broadening of the FM transition and the decrease in $1/T_1T$, as shown at the bottom of Fig. \ref{Fig:T1T-HcDep}. 
Fortunately, we can use $1/T_1T$ measured by the NQR technique as $1/T_1T$ under zero external field, since the principal axis of the electric field gradient is almost perpendicular to the $c$ axis, and $1/T_1T$ by NQR mainly detects the magnetic fluctuations along the $c$-axis, which is the same as $1/T_1T$ measured in $H~||~b$. 
However, the shift of the peak temperature from the NQR data, which is actually observed at $\mu_0H \sim$ 3 T parallel to the $b$-axis, suggests a very small misalignment (less than about 0.5$^\circ$). 
Another possibility is that this small shift of \TCurie~might be an intrinsic phenomenon, since a small increase in \TCurie~by applying the magnetic field along the $b$-axis was reported from the macroscopic measurements\cite{DAoki2009JPSJ}, 
We measured $1/T_1T$ up to $\mu_0 H \sim 12$ T without changing the angle condition once the angle was set along the $a$-and $b$-axes.

\begin{figure}[htbp]
 \includegraphics[width=\hsize,clip]{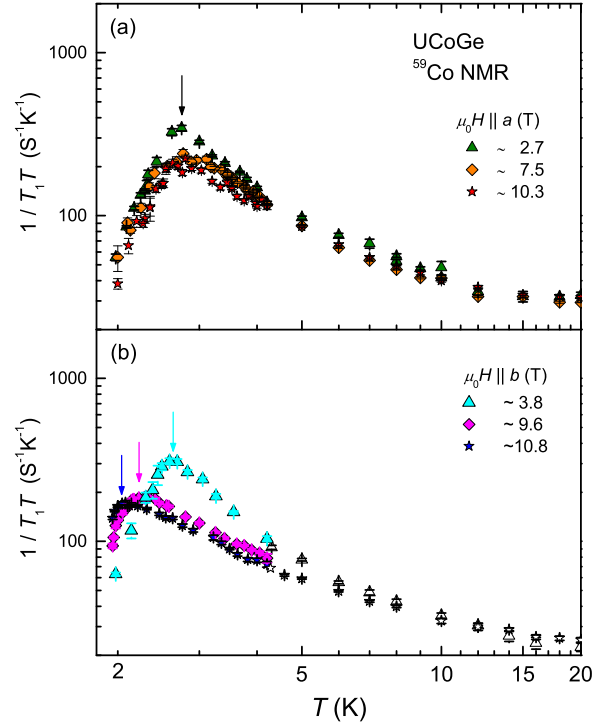}
 \vspace*{-10pt}
 \caption{(color online) Temperature dependence of \Co~NMR $1/T_1T$ with various fields along the ($a$) $a$-axis  and ($b$) $b$ axis . \Co~NMR $1/T_1T$s below 4.2 K were measured with the center and 3rd satellite NMR peaks. Qualitatively, both show the same behavior, and the peak temperature does not depend on the NMR site. Temperature dependences of $1/T_1T$ above 4.2 Kelvin were measured with the center peak.}
  \label{Fig:1oT1T}
\end{figure}

\begin{figure}[htbp]
 \includegraphics[width=\hsize,clip]{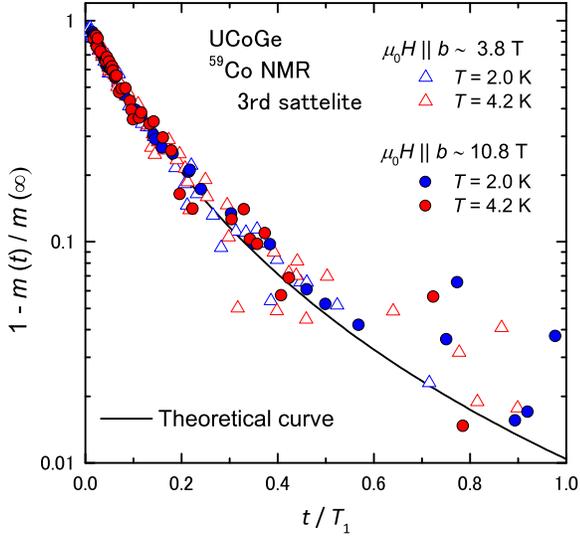}
 \vspace*{-10pt}
 \caption{(color online) Recovery curves $R(t)$ of the nuclear magnetization $m(t)$ measured at the \Co~3rd satellite NMR peaks ($E_{7/2} \Leftrightarrow E_{5/2}$) at $t$ after a saturation pulse. The $x$-axis is normalized by $T_1$ and the line presents the theoretical function for the $T_1$ evaluation.}
  \label{Fig:Rec-Normalized}
\end{figure}

Figure \ref{Fig:1oT1T} shows the temperature dependence of $1/T_1T$ up to a high field along the $a$- and $b$-axes. 
The $1/T_1T$ below 4.2 K was measured at the 3rd satellite peak of the \Co~NMR spectrum, which corresponds to the $E_{7/2} \Leftrightarrow E_{5/2}$ transition in the \Co~nuclear spin levels. 
The temperature dependence of $1/T_1T$ above 4.2 K was measured at the central peak corresponding to the $E_{1/2} \Leftrightarrow E_{-1/2}$ transition, and both data are consistent with each other. 
The recovery curves $R(t)$ of the nuclear magnetization $m(t)$ measured at the 3rd satellite peaks at 2.0 and 4.2 K, in which the $x$ axis values are normalized by $T_1$ values, are shown in Fig. \ref{Fig:Rec-Normalized}. 
In the figure, we show the results for $\mu_0 H~||~b \sim 3.8$ and 10.8 T as examples.
The consistent fitting of $R(t)$ by the theoretical function (black line in Fig. \ref{Fig:Rec-Normalized}) indicates that the electronic state in UCoGe around \TCurie~is homogeneous, regardless of field values. 
Therefore, we can confirm that the different response of $1/T_1T$ against the field direction is purely due to the change in the magnetic fluctuations along the $c$-axis. 

\begin{figure}[htbp]
 \includegraphics[width=\hsize,clip]{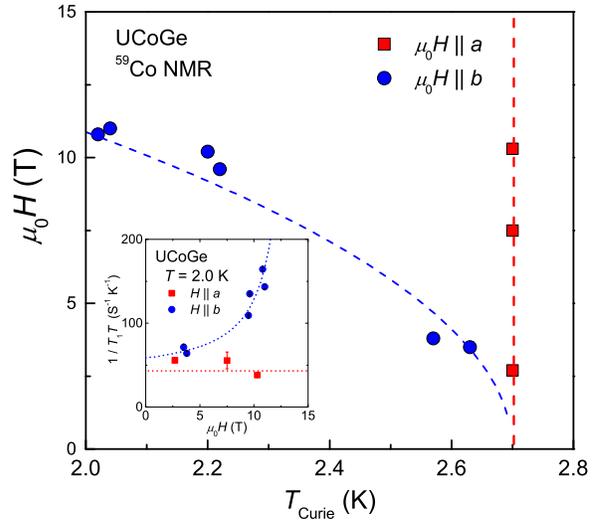}
 \vspace*{-10pt}
  \caption{(color online) Field dependence of \TCurie~determined by the peak of $1/T_1T$ against temperature in the fields along the $a$- and $b$-axes. The blue line in the main panel represents the relation: $T_{\rm Curie} (H^{\perp}) = T_{\rm Curie} -\alpha (H^{\perp})^2$. 
The inset shows the field dependences of $1/T_1T$ measured by \Co~NMR at $T =$ 2.0 K ($< \TCurie$). The dotted lines in the inset are guides to the eye. While the field along the $a$-axis does not change the magnetic properties, the magnetic field along the $b$-axis enhances the magnetic fluctuations.}
\label{Fig:HDep-TCurie}
\end{figure}

Now, we discuss the field dependences of $1/T_1T$ and  \TCurie~behavior. 
In general, when a magnetic field is applied to ferromagnets, FM transition becomes blurred. 
However, a clear anomaly can be detected in UCoGe under the external field perpendicular to the Ising spontaneous moment. 
Therefore, we can unambiguously determine \TCurie~even under the magnetic field, which is shown in Fig. \ref{Fig:1oT1T} by the arrows. 
Although \TCurie~hardly changes in $H~||~a$, \TCurie~decreases in $H~||~b$, and this \TCurie~decrease is in good agreement with the previous result determined from the kink of the electric resistivity under the field \cite{DAoki2009JPSJ}, as shown in Fig. \ref{Fig:HDep-TCurie}.  
This decrease in \TCurie~ is related to the enhancement of the FM fluctuations at low temperatures, which is shown by the field dependence of $1/T_1T$ at 2.0 K (inset of Fig.~4).
At the moment, we cannot exclude the possibility that magnetic fluctuations along the $b$ axis are induced by $H~||~b$, but it is natural to consider that the Ising FM fluctuations along the $c$-axis, which are dominant at low fields, are enhanced with the ordered moments pointing to the $c$-axis, since the spin-flop behavior has not been observed in the field dependence of the magnetization at $\mu_0H^b \sim 12$ T below \TCurie~in UCoGe \cite{WKnafo2012PRB}. 
To probe the spin fluctuations along the $b$-axis, it is important to measure the nuclear spin-spin relaxation rate $1/T_2$ in the critical field range of $H^b$, since $1/T_2$ can probe the spin fluctuations along the applied-field direction, which is now in progress.    

From the theoretical viewpoint, Mineev discussed the field dependence of the effective amplitude of SC pairing interaction as well as \TCurie~and the $c$-axis magnetization\cite{VPMineev2011PRB}.
In his model, the orbital depairing effect is not taken into account.
In what follows, we analyze the experimental results on the basis of Mineev's discussion. 
When $H$ is applied perpendicular to the $c$-axis [$H^{i}$ ($i$ = $a$- or $b$-axis)], \TCurie~ and the magnetization along the $c$-axis ($M_c$) are suppressed, and the low-temperature susceptibility along the $c$ axis is enhanced with $H$ by following the relation of 
\begin{eqnarray}
\alpha_i H_{i}^{2}&=&T_{\rm Curie} (H^i=0)- T_{\rm Curie}(H^i)\\
                     &\propto& M_c(H^i = 0)^2 - M_c(H^i)^2\\
                     &=&\frac{1}{\chi_c(H^i=0)}-\frac{1}{\chi_c(H^i)}  .
\end{eqnarray}
Here, $\alpha_i$ is the coupling constant between $M_c$ and $H^{i}$. 
These relations are derived from the Landau free energy of the orthorhombic ferromagnet in the magnetic field.
The \TCurie~suppression roughly following this relation was observed in $H^b$, and is shown in Fig. \ref{Fig:HDep-TCurie} by the blue dotted line.
In addition, $1/T_1T$ at $T=2$ K (the lowest temperature in the present measurements), which is related to the susceptibility along the $c$-axis, is enhanced as $H^b$ increases (Fig.~\ref{Fig:HDep-TCurie} inset), as a result of decreasing $T_{\rm Curie}$. 
In contrast, the field parallel to the $a$-axis does not suppress \TCurie~and $1/T_1T$ at  2 K is unchanged up to 11 T, indicative of $\alpha_a$ being negligibly small.  
This different response against the applied field would give an important clue to understanding the anisotropy of \Hc~along the $a$- and $b$-axes.
The increase in $1/T_1T$, as well as the effective mass\cite{DAoki2009JPSJ} in $H^b$, suggests that the Ising FM fluctuations just above the SC transition temperature ($\TSC \sim 0.6$ K) are enhanced with the magnetic field along the $b$-axis ( $\mu_0 H~||~b \sim 11 $ T), while the FM fluctuations around \TSC~are unchanged with $H^a$.

\begin{figure}[htbp]
 \includegraphics[width=\hsize,clip]{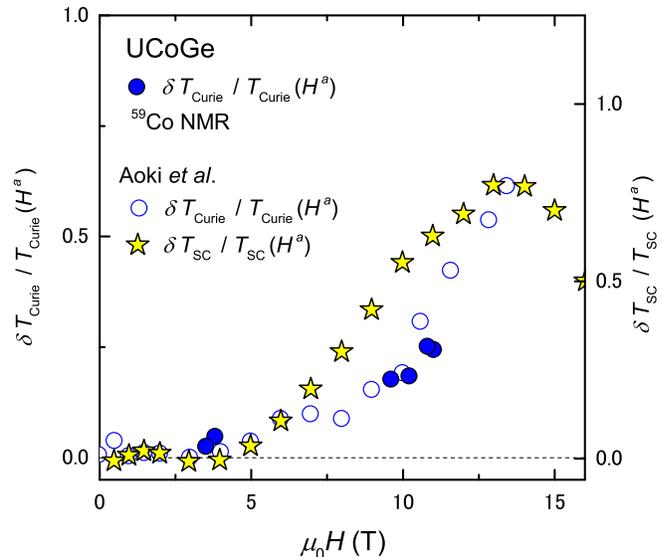}
 \vspace*{-10pt}
  \caption{(color online) Plot of the difference in \TCurie~in $H^a$ and $H^b$ [$\delta T_{\rm Curie} \equiv T_{\rm Curie}(H^a) -T_{\rm Curie}(H^b)$] against the magnetic field. Here, the unchanged \TCurie in $H^a$ is assumed to continue up to 16 T. The star points show the deviation of superconducting transition temperature $\delta \TSC = \TSC (H^b) - \TSC (H^a)$ when $H$ is applied parallel to the $a$- and $b$-axes. The $\delta$\TCurie shown by the open circles and $\delta T_{SC}$ shown by stars are referred to from the reference by D. Aoki {\it et al.}\cite{DAoki2009JPSJ}. }
\label{Fig:Fig5}
\vspace*{-10pt}
\end{figure}

Therefore, it is meaningful to compare the field-induced spin susceptibility along the $c$-axis with the anisotropy of superconductivity using the experimental results obtained along the $a$- and $b$-axes\cite{DAoki2009JPSJ}, since the SC pairing interactions are predicted  to be coupled to the spin susceptibility in the scenario of the spin-fluctuation-mediated superconductivity\cite{DFay1980,PMonthoux1999PRB,VPMineev2011PRB}.
We assume that the unchanged \TCurie~in $H^a$ continues up to 16 T, and plot the difference in \TCurie~in $H^a$ and $H^b$ [$\delta T_{\rm Curie} \equiv T_{\rm Curie}(H^a) -T_{\rm Curie}(H^b) >0$] against the magnetic field in Fig. \ref{Fig:Fig5}.
In the figure, the \TCurie~ determined from the resistivity is also plotted\cite{DAoki2009JPSJ}. 
From the above equations, it is shown that $\delta T_{\rm Curie}$ is related to the enhancement of the $c$-axis susceptibility induced by $H^b$ with the relation $\delta T_{\rm Curie} = 1/\chi_c(H^a)- 1/\chi_c(H^b)$. 
In the same figure, if we plot the deviation of the SC transition temperature [$\delta T_{\rm SC} \equiv T_{\rm SC}(H^b) -T_{\rm SC}(H^a)$] in $H$ perpendicular to the $c$-axis, as reported by Aoki {\it et al}.\cite{DAoki2009JPSJ}, we notice that $\delta T_{\rm Curie} $ and $\delta T_{\rm SC}$ behave in nearly the same manner against magnetic fields.
This is a clear indication that the robustness of the superconductivity observed in $\mu_0 H^b$ from 5 to 15 T is related to the enhancement of the Ising FM fluctuations by $H^b$, since $\chi_c(H^b) > \chi_c(H^a) \sim \chi_c(H~=~0)$ is shown from $\delta T_{\rm Curie} > 0$.
Although the orbital depairing effect is not taken into account in the present discussion, the relatively good scaling between $\delta T_{\rm Curie}$ and $\delta T_{\rm SC}$ suggests the validity of the scenario of the spin-fluctuation-mediated superconductivity in UCoGe.   
By taking into account the responses of the Ising FM fluctuations against magnetic fields along the three crystalline axes, we strongly suggest that the Ising FM fluctuations tuned by magnetic fields play an important role in superconductivity as a glue of the Cooper pairs, accompanied by the suppression of the pair breaking effect owing to the increase in the effective mass. 
In the present case, since the superconductivity is induced by the FM fluctuations, the pairing state is expected to be a spin-triplet state\cite{DFay1980,PMonthoux1999PRB,TRKirkpatrick2001PRL,ZWang2001PRL,RRoussev2001PRB,AVChubukov2003PRL,SFujimoto2004JPSJ}. 
In fact, the spin-triplet state with the spin component along the $c$-axis was suggested from the Knight-shift measurement in the field along the $a$- and $b$-axes \cite{THattori2013PRB}.

Here, we discuss the similarity and difference in the SC enhancement observed between URhGe and UCoGe. 
Both compounds show the FM ordering with the Ising character along the $c$-axis (ordered moment: $m_c \sim 0.40~\mu_{\rm B}$ in URhGe and $m_c \sim 0.05~\mu_{\rm B}$ in UCoGe \cite{DAokiJPSJ2012}. Both are much smaller than the Curie term estimated from the bulk susceptibility above $T_{\rm Curie}$). 
In URhGe, the re-entrant superconductivity was observed at $\mu_0 H^b \sim 11$ T, where the spin-flop anomaly was observed in the $b$-axis magnetization \cite{FLevy2005URhGeReSC,FLevy2007URhGePD}. 
Although the recent angle-resolved photoemission spectroscopy (ARPES) measurement strongly suggests that the itinerant description of U-5$f$ states is a good starting point for understanding the magnetism on URhGe \cite{SFujimori2014PRB}, the reentrant superconductivity seems to be related to the localized character of the U-5$f$ states: the soft magnon induced by spin orientation to the $b$-axis is suggested to generate the SC pairing interactions \cite{KHattori2013PRBURhGe}. 
To check this scenario, NMR measurements on a high-quality single-crystal URhGe are highly desired. 
On the other hand, since the localized moment is small and such spin-flop has not been observed at around the SC robustness in UCoGe as discussed above \cite{WKnafo2012PRB}, the itinerant character rather than the localized one in the U-$5f$ electrons would be important for the SC robustness in $H^{b}$. 
In addition, we speculate that the SC symmetry and mechanism would be the same in the whole SC region of UCoGe from the absence of separation of the SC region against $H^b$, although it should be checked in future measurement. 

Finally, we comment on the origin of the FM critical behavior induced by $H^b$ in UCoGe.  
As mentioned above, the magnetism on UCoGe should be considered on the basis of the itinerant nature of the U-$5f$ electrons, therefore the Fermi-surface (FS) properties are important for understanding the magnetism. 
Recently, the modification of the FS at around  $\mu_0 H = 11$ T along the $b$-axis has been suggested from the thermopower measurement on UCoGe \cite{LMalone2012PRBUCoGeFS}. 
We consider that such modifications would give a strong effect on the magnetic properties, and that the correlation between the FS modification and the enhanced superconductivity can be consistently understood if the superconductivity is induced by the ferromagnetic fluctuations originating from the FS nesting.

 In conclusion, we measured $1/T_1$ of \Co~under the external field perpendicular to the $c$-axis. 
We found that \TCurie~is unchanged with $H^a$ up to 11 T, but \TCurie~decreases with increasing $H^b$, resulting in the longitudinal FM fluctuations at $T = 2$ K being unchanged under $H^a$, but they are enhanced under $H^b$ greater than 5 T, where the enhancement of superconductivity was reported \cite{DAoki2009JPSJ}. 
From the field dependences of $\delta T_{\rm Curie}$ and $\delta T_{\rm SC}$, we show that the robustness of superconductivity observed in $\mu_0 H^b$ from 5 to 15 T is strongly related to the enhancement of the Ising FM fluctuations by $H^b$, which is indicative of the validity of the scenario of the spin-fluctuation-mediated superconductivity.   
This might be another piece of evidence that the Ising FM fluctuations induce the superconductivity in UCoGe.

\section*{Acknowledgments}
The authors thank Y. Ihara, S. Yonezawa, and Y. Maeno for experimental support and valuable discussions, and Y. Tada, S. Fujimoto, H. Ikeda, A. de Visser, D. Aoki, V. P. Mineev and J. Flouquet for valuable discussions. This work was partially supported by the Kyoto Univ. LTM Center, Yukawa Institute, the ``Heavy Electrons" Grant-in-Aid for Scientific Research on Innovative Areas  (Nos. 20102006, 21102510,  20102008, and 23102714) from The Ministry of Education, Culture, Sports, Science, and Technology (MEXT) of Japan, a Grant-in-Aid for the Global COE Program ``The Next Generation of Physics, Spun from Universality and Emergence'' from MEXT of Japan, and a Grant-in-Aid for Scientific Research from the Japan Society for the Promotion of Science (JSPS), KAKENHI (S) (No. 20224015), the Inter-University Cooperative Research Program of the Institute for Materials Research, Tohoku University (Proposal No. 14K0068), and JSPS Research Fellow.

\bibliographystyle{jpsj2.bst}
\bibliography{16928Ref.bib}
\end{document}